# Evaluation of the Effects of Frame Time Variation on VR Task Performance

Benjamin Watson, Victoria Spaulding, Neff Walker, and William Ribarsky
Graphics, Visualization, and Usability Center
Georgia Institute of Technology

**Abstract**

*We present a first study of the effects of frame time variations, in both deviation around mean frame times and period of fluctuation, on task performance in a virtual environment (VE). Chosen are open and closed loop tasks that are typical for current applications or likely to be prominent in future ones. The results show that at frame times in the range deemed acceptable for many applications, fairly large deviations in amplitude over a fairly wide range of periods do not significantly affect task performance. However, at a frame time often considered a minimum for immersive VR, frame time variations do produce significant effects on closed loop task performance. The results will be of use to designers of VEs and immersive applications, who often must control frame time variations due to large fluctuations of complexity (graphical and otherwise) in the VE.*

## 1 Background and motivation

There have been many studies on the effects of frame update rates in immersive virtual environments on task performance, the sense of presence, the propensity for motion sickness, and other factors. These studies choose target frame rates that are held (or assumed to be) constant during the course of the experiments.

It is also often assumed that frame rates *should* be held constant to ensure the best performance in the virtual environment. Indeed, significant effort has been expended recently to come up with techniques that ensure constant or near constant frame rates [5,6,9] even as the amount of detail varies greatly from scene to scene. These studies establish a metric, usually in terms of polygonal count, that can be adjusted to speed up or slow down frame update rate. In addition adaptive detail management systems [5,6] provide a mechanism for adjusting the per object polygon count as the user moves through an environment encountering varying numbers of objects. The overall effect is to achieve a more or less constant number of total polygons in each scene. However, if the adaptation is achieved entirely by feedback (the next frame metric being adjusted based on the timing of the previous frame), there will tend to be an overshooting and oscillation in frame rate, especially when there is an abrupt change in detail (as when the user turns a corner from an empty room to one filled with objects). Funkhouser and Sequin [5] have shown that a predictive method can overcome this problem for architectural walkthroughs. In principle their approach is general; however, it has not been implemented for other cases. Certainly, there can easily be more complicated situations than the one they considered--for example, ones with lots of rapidly moving objects, or multiresolution terrain plus architectural elements, significant simulations launched as a result of user actions, and so on. For these cases, it is not clear how exactly to go about setting up a completely predictive model and how successfully the model will control frame rate variations (especially since constraining by minimizing time costs while maximizing scene benefits is an NP-complete problem). Faced with these difficulties and a choice of methods (e.g., feedback versus predictive), it would be good if an application designer had some idea of the likely effects that frame rate variation, as a function of average frame rate, has on her application tasks.

Lag, the time delay between a user action and its displayed result, is intimately connected with frame rate and must also be considered by application designers. As Wloka has pointed out [15], there will be a *synchronization lag*, on top of any other sources of lag, that will vary from zero to the whole frame time (e.g., 100 ms for a frame rate of 10 fps) depending on when in the frame cycle user input is collected. Thus there will always be a variation in lag that will grow more pronounced as frame rate variations grow.

There is little experimental work so far that would help designers to factor frame rate and lag into their design decisions. Early research in the field of teleoperation [4,12] focused on lags above a second. Mackenzie and Ware [10] considered lag in 2D displays. Ware and Balakrishnan [14] varied lag and frame rate in a fishtank (head-tracked, stereoscopic, table-top) display. Performance on a placement task declined as task difficulty and hand-tracked lag increased. Lag in head tracking had little effect (probably due to the type of display), and lags were not varied during the task itself.

In this paper we present experimental results for generic grabbing and placement tasks in a VE with head-mounted display. These tasks are of the type that are often required in VR applications and thus provide a significant starting point for filling in the VE design space. Using a set of variations in both average frame rate and deviation around the average, we measured both accuracy and time for performing these tasks. We do not separate the effects of lag and frame rate in these experiments. The experimental results allow us to draw some conclusions about frame rates and their variations and suggest further studies.

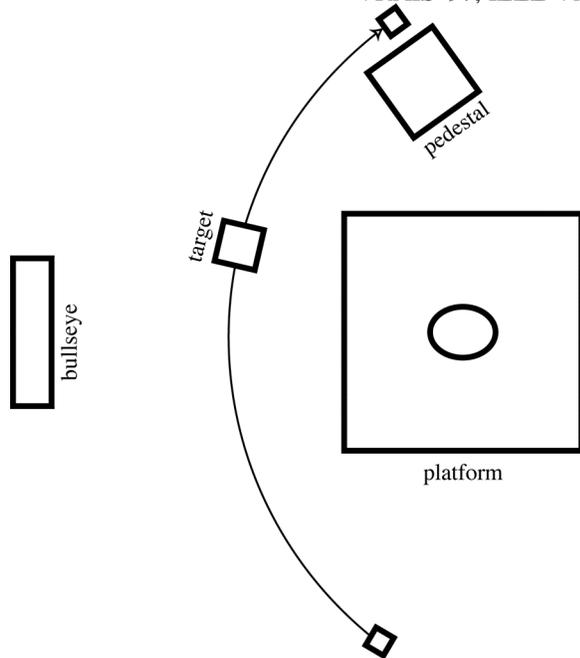

Figure 1: A top down schematic of the experimental environment. Users on the platform begin by looking at the bullseye; the target moves left to right across the visual field.

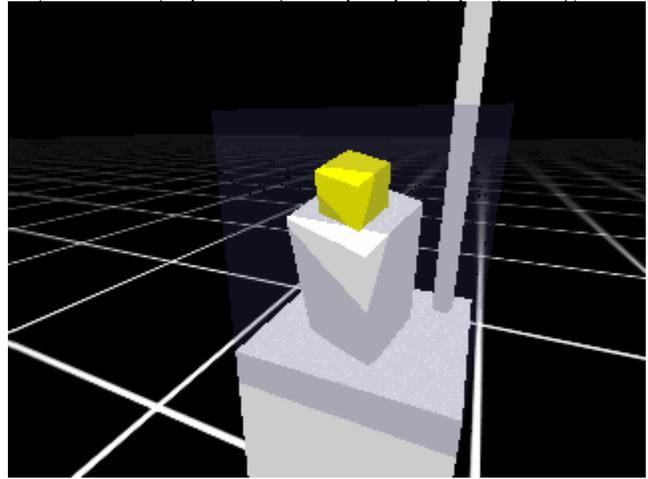

Figure 2: View after a trial with unsuccessful placement. The large white object at the bottom is the pedestal, the target is immediately on top of it. The cursor is on top of the target. The target leans past the front edge of the pedestal and through the translucent placement box, a common mistake.

## 2 Experimental setup

### 2.1 Participants

There were 10 participants in the study, a mixture of undergraduate and graduate students. These were both somewhat experienced (graduate) and inexperienced (undergraduate) users of virtual reality and head-mounted displays. Although one or two of the inexperienced participants had lower performance scores than the others, there was no overall trend in performance ranking based on experience. Vision for all participants was normal or corrected-to-normal (via contact lenses). The subject with the best cumulative ranking at the end of the experiment received 50 dollars. Undergraduate subjects also received credit in an introductory course for their participation.

### 2.2 Apparatus

The virtual environment was displayed with a Virtual Research VR4 head-mounted display, with a 36º vertical field of view and a 48º horizontal field of view. The two screens in the display overlap fully and each contains 247 x 230 color triads with resolution of 11.66 arcmin. The display was used in a biocular mode, with the same image shown to each eye. Head position was tracked with a Polhemus Isotrak 3D tracker, with an effective tracking radius of approximately 1.5 M. A Crimson Reality Engine generated the images. The subjects interacted with the environment using a plastic mouse, shaped like a pistol grip. During the experiment, they stood within a 1 M X 1 M railed platform. The platform was 15 cm high and the railing was 1.12 M high.

### 2.3 The task

The participants visually tracked a moving target object, grabbed it, and placed it on a pedestal within a certain tolerance. The target object was a white oblong box measuring 0.31 M in height and 0.155 M in depth and width. If the participant stood in the center of the platform, the target flew by on an arc of constant radius 0.69 M that subtended an angle of 125º. The pedestal was at one end of the arc (0.69 M away). (See Fig. 1.) Thus the target and pedestal could be reached without leaning by an average-sized person. A small, yellow cubic cursor, 0.09 M across each side, represented the joystick/hand location within the virtual environment. Visual cueing guided the subject's grasp of the object: the target turned yellow and the cursor white when the subject successfully grasped the target.

The virtual environment consisted of a black floor with a white grid superimposed on it, and a black background. The ends of the target's arc of motion were marked by tall white posts (as shown in Fig. 1). After reaching the end of the arc and after a 1.5 second pause, the target reappeared at the left of the arc, giving the effect of a wraparound motion. The target moved up and down in a sinusoidal pattern. The amplitude of the sine wave measured 0.85 M, and the target described a complete period of the sine wave after traveling across the arc. The phase of the sine was chosen randomly each time the target appeared at the left end of the arc.

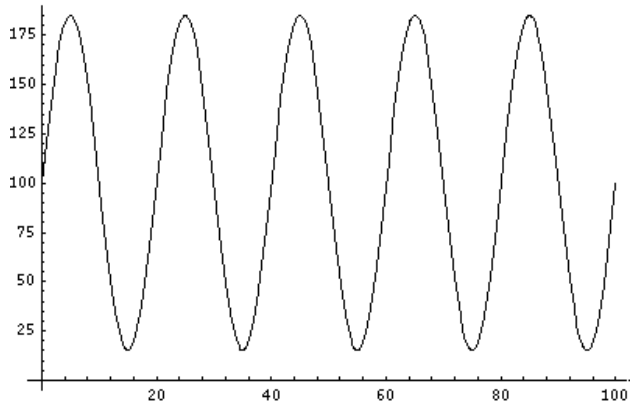

Figure 3a: A plot of targeted frame time (ms) versus frame number for the 100 ms mean, 60 ms deviation, 20 frame fluctuation period condition.

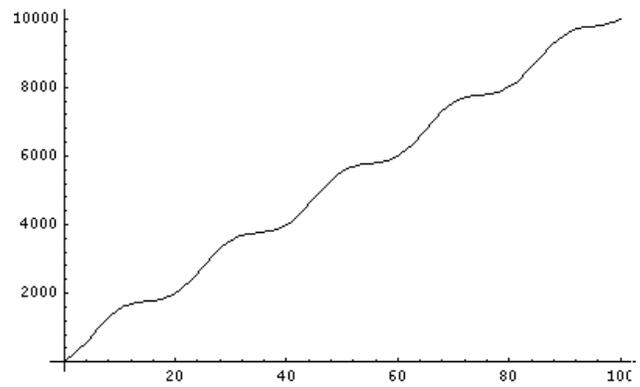

Figure 3b: A plot of targeted elapsed time (ms) versus frame number for the 100 ms mean, 60 ms deviation, 20 frame fluctuation period condition.

The pedestal was white and located next to the base of the post marking the right end of the arc. It was an oblong box 1.5 M tall, and 0.45 M in depth and width. We settled on this depth and width for the pedestal after some trial and error, making the area large enough so that placement could be accomplished without excessive attempts. Success of the placement task was measured by testing the location of the target object: it had to be completely contained in a placement box. The placement box had the same depth and width as the pedestal, and measured 55 cm in height. It was blue and transparent, and was only visible as feedback after the target was incorrectly placed. A typical incorrect placement is shown in Fig. 2 where the base of the object is within the placement box but the top end is tilted out.

To ensure uniform trials, participants could not begin a trial until they centered a red and white bullseye in their view. The bullseye was centrally positioned on a solid black background between the two posts.

The tasks that the participants had to accomplish were of two different types. The grabbing of the moving target was mostly an open loop task while the placement on the pedestal was a closed loop task. Open loop tasks involve movements that do not allow feedback or correction, such as throwing a ball at a target. Once the movement has been planned and made, no course corrections can be made. A closed loop task is one in which a person makes an initial movement, then gets feedback about the correctness of the movement, and makes further movements to correct for error. Because of their different strategies of movement, these tasks may be affected differently by frame time fluctuations. Both tasks fall into the VE performance assessment battery set up to compare task performance across VE systems [8]. In battery, the grabbing and placement tasks are manipulation tasks.

## 3 Frame rate and lag variations

As soon as one plans an experiment that studies frame rate variation (and the concomitant variation in lag), one must consider both the amplitude of the deviation and its frequency. Frame rate is an average quantity, so it seems better to consider variations in the directly measured quantity, frame time (the length of each frame), as a function of the number of frames. We can then always find an average frame rate over a time period by dividing the number of frames passed during the period by the time. Since we ensured that the system would run well above the target frame times, we can reach the target by adding an appropriate delay time at each frame. Actual frame times/lags were recorded to confirm this experimental control. Each frame was rendered in the following loop: first, tracker location was obtained, next delay was added, and third, the frame was rendered. By adding delay in this fashion, we caused lag to vary with the same frequency as the frame time. As an alternative we could have removed this lag variation by adding the delay *after* rendering of the frame and before obtaining the new tracker position. If the tracker updates and frame rendering are fast with respect to the target frame time, the differences in frame time fluctuation between the two methods will be small. End-to-end lag in our system without delay averages 213 ms with a 30 ms standard deviation. With delays, the average lag is 235 ms for the 50 ms frame time and 285 ms for the 100 ms frame time.

We decided to impose frame time variations in a simple, controllable way by using a sinusoidal variation versus frame number as shown in Fig. 3a. The period of the sine wave gives the frequency of oscillation. By integrating under the curve, we can find elapsed time versus frame number as shown in Fig. 3b. It is now easy to follow this curve by merely adding a delay at each frame to make the accumulated time match the calculated elapsed time. Our measurements confirm that we can achieve the appropriate average frame time and the detailed fluctuation behavior using this method.

## 4 Experimental design and procedure

### 4.1 Design

The study used a 2 (mean frame time) X 3 (fluctuation amplitude) X 2 (period of fluctuation) design. Thus there were 12 display conditions determined by the three independent variables. The mean frame times were 100 ms and 50 ms, which are 10 fps and 20 fps, respectively. This frame rate range brackets many VR applications. Several researchers consider 10 fps a minimum for immersive virtual environments [3,10]. For fully "acceptable" performance, higher frame rates are often required, such as 10-15 fps [11] in a survey of display systems and their characteristics, at least 15 fps for military flythroughs [7], and up to 20 fps for certain architectural walkthroughs [1]. There were three fluctuation amplitudes with standard deviations of 20 ms, 40 ms, and 60 ms for the 100 ms mean frame time and 10, 15, and 20 ms for the 50 ms mean frame time. Finally, the two periods for the sine wave oscillation were 5 frames and 20 frames. All these conditions are summarized in Tables 1 and 2.

The reason for choosing two different sets of fluctuation amplitude standard deviations is that otherwise one runs into trouble with the larger deviations. If we were to use the same deviation values in both cases, obviously a deviation of 60 ms would not work for a 50 ms frame time. An alternative is to use the same percentages. Here 60 ms is 60% of the 100 ms frame time, so the corresponding deviation at 50 ms is 30 ms. However, the latter gives a range of frame times whose low standard deviation is 20 ms (50 fps, with actual lowest frame time of 50 - sqrt(2) x 30 = 8 ms), which we cannot consistently reach on our Crimson Reality Engine with the present virtual environment. We decided to forgo any direct matching of fluctuation standard deviations in favor of covering the range where there were likely to be significant effects at each frame time. This makes detailed comparisons between frame times harder, but this problem can be alleviated, if desired, by filling in with more trials at additional fluctuation amplitudes.

There were 5 dependent measures, 3 for time and 2 for accuracy. The time measures were mean grab time (average time to successfully grab the target), mean placement time (average time to successfully place the target on the pedestal), and mean total time (average time to complete a trial). These mean times were calculated for the correct trials. The measures of accuracy were percentage of trials correctly performed and the mean number of attempts to grab the target.

### 4.2 Procedure

Each person participated in two sessions. Each session consisted of one block of twenty practice trials, followed by twelve blocks of experimental trials. One display condition was presented in each experimental block. Three practice trials were presented at the onset of each display condition. Accurate placement of the target within thirty seconds was defined as a correct trial, and each subject had to complete five correct trials per block, for a total of 120 correct trials per subject over both sessions. Incorrect trials were discarded and subjects were required to complete all trials within each display condition before ending the session. The presentation order of the blocks was varied randomly between subjects and each order was used once.

A trial consisted of the subject orienting on the bullseye location and pressing the trigger button on the joystick to begin. After a random delay (between 750 ms and 1750 ms) the target appeared, and the bullseye disappeared. The target moved at a fixed horizontal velocity of 0.75 m/sec and followed the sinusoidal path described in Sec. 2. To grab the target, the subject had to press the trigger button while the yellow cursor intersected the target. When the target had been successfully grabbed, it would shift to a location underneath the cursor. This made placement difficulty independent of grasp location. To complete the trial, the subject moved the target to the right side of the visual field and placed it on the pedestal. For the placement to be correct, the target rectangle had to be placed completely inside the placement box as described in Sec. 2.

## 5 Results

The data were analyzed by means of five three-way analyses of variance (mean frame rate by fluctuation amplitude by period of fluctuation). The analyses were performed on mean grab time, mean positioning time, mean total time, mean number of grab attempts, and percent correct trials. The means of times were based on correct trials only. Bonferroni pair-wise comparisons and simple main effects tests were used to follow up any significant effects. In order to save space, only effects that reached at least a marginal level of significance ($p < 0.10$) will be reported. Results are shown for all twelve conditions in Tables 1 and 2. Grab and placement times are graphed in Figures 4a and 4b.

The main significant effect in the experiment occurred for the placement time (to put the target on the pedestal) at mean frame time of 100 ms. Frame time fluctuation and the period of fluctuation interacted significantly ($p = 0.04$) for the placement time. When the fluctuation amplitude was less than 60 ms, placement times were similar for both the 5 frame and 20 frame periods. However, at the 60 ms fluctuation, the period had a significant effect, resulting in very dissimilar placement times: 2.20 sec at 5 frames versus 3.04 sec. at 20 frames. At 20 frames, the 60 ms result was significantly larger than those at lower

Table 1: Display conditions and results for a frame time of 100 ms. Mean lag in these conditions was 235 ms. Since the time distribution is a sine wave, the frame time range = 1.414 * (standard deviation); e.g., for a standard deviation of +/- 20 ms, the frame range is 72 - 128 ms. Lag varied in a similar fashion.

| Conditions | | | Results | | | | |
| --- | --- | --- | --- | --- | --- | --- | --- |
| Period Frame Time Change | Std Dev Frame Time | *Rng Of Frame Time (Rate)* | Avg Num Grabs | Pct Trials Correct | Avg Grab Time | Avg Place Time | Avg Total Time |
| 5 | 20 | *72-128 (7.8-13.9)* | 1.62 | 97.4 | 2.64 | 2.38 | 5.02 |
| 5 | 40 | *43-157 (6.4-23)* | 1.61 | 89.7 | 2.71 | 2.50 | 5.21 |
| 5 | 60 | *15-185 (5.4-66)* | 1.66 | 88.7 | 2.85 | 2.21 | 5.06 |
| 20 | 20 | *72-128 (7.8-13.9)* | 1.97 | 84.2 | 3.88 | 2.41 | 6.29 |
| 20 | 40 | *43-157 (6.4-23)* | 1.64 | 88.2 | 2.98 | 2.45 | 5.43 |
| 20 | 60 | *15-185 (5.4-66)* | 2.01 | 85.3 | 4.05 | 3.04 | 7.09 |

Table 2: Display conditions and results for a frame time of 50 ms. Mean lag in these conditions was 285 ms.

| Conditions | | | Results | | | | |
| --- | --- | --- | --- | --- | --- | --- | --- |
| Period Frame Time Change | Std Dev Frame Time | *Rng Of Frame Time (Rate)* | Avg Num Grabs | Pct Trials Correct | Avg Grab Time | Avg Place Time | Avg Total Time |
| 5 | 10 | *36-64 (15.6-27.9)* | 1.38 | 92.9 | 2.17 | 2.13 | 4.30 |
| 5 | 15 | *29-71 (14-34.7)* | 1.51 | 93.6 | 2.25 | 1.81 | 4.06 |
| 5 | 20 | *22-78 (12.8-46)* | 1.47 | 93.0 | 2.21 | 2.03 | 4.24 |
| 20 | 10 | *36-64 (15.6-27.9)* | 1.34 | 93.5 | 2.02 | 2.09 | 4.11 |
| 20 | 15 | *29-71 (14-34.7)* | 1.35 | 90.7 | 2.12 | 1.94 | 4.06 |
| 20 | 20 | *22-78 (12.8-46)* | 1.30 | 91.5 | 1.89 | 2.04 | 3.93 |

fluctuation amplitudes whereas the 5 frame results were not significantly different.

The percent of successful trials and the number of grab attempts per trial were not significantly affected by changes in display variables for the 100 ms frame time. However, the effect of fluctuation period on grab times was marginally significant (p = 0.09) with average grab times (over all fluctuation amplitude deviations) being 2.74 sec at 5 frames versus 3.64 sec at 20 frames. In addition the period also showed a marginally significant effect (p = 0.08) on the total time, with average times over all fluctuation deviations of 5.1 sec at 5 frames versus 6.27 sec at 20 frames.

The 50 ms frame time trials were run with fluctuation standard deviations of 10, 15, and 20 ms at periods of 5 and 20 frames. There were no significant dependencies on these display variables for any of the dependent measures (grab time, placement time, total time, number of grab attempts, and percentage of trials correctly performed). However, changes in the period had a marginally significant effect (p = 0.098) on the number of grab attempts, while changes in fluctuation deviation marginally affected (p = 0.07) the placement time.

Although we cannot compare them in detail because of the different fluctuation standard deviations used, it is interesting to note in general terms the differences between results at 100 ms and 50 ms frame times. The average placement times and grab times were 2.50 and 3.20 sec at 100 ms versus 2.01 and 2.11 sec at 50 ms. The average number of grab attempts and percentage of correct trials were 1.75 and 0.89 at 100 ms versus 1.36 and 0.93 at 50 ms. Clearly user performance improves significantly in going from 100 ms to 50 ms average frame time for all dependent measures. This result for the open and closed loop tasks in this experiment is consistent with results on other tasks and applications [3,11,15].

## 6 Discussion

A main conclusion from this study is that at low enough frame times (certainly by 50 ms or 20 fps) symmetrical changes in fluctuation amplitude (at least up to 40% about the mean) and changes in fluctuation period have little or no effect on user performance for the two types of tasks presented here. Further at frame times high enough (certainly by 100 ms or 10 fps), not only is general performance of tasks in terms of time and accuracy degraded, but performance can depend on both fluctuation amplitude deviation and fluctuation period. A general conclusion is that if average frame rate is high enough, a VR application designer need not worry so much about retaining tight control over fluctuations around the mean. Further, when prediction of performance is necessary, it will require taking into account details of the frame rate variation over time if the average frame time is high enough (or average frame rate low enough).

We further see differences between the mostly open loop task (grabbing) and the closed loop task (placement)

VRAIS '97, IEEE Virtual Reality Annual Symposium (Albuquerque, April, 1997), 38-44

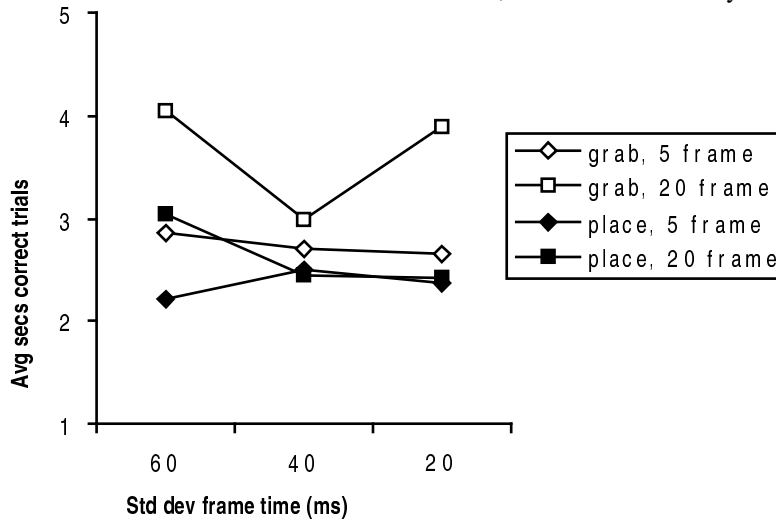

Figure 4a: Mean grab and place times for the mean frame time of 100 ms, with task type indicated by point color and frame time frequency by shape.

in the experiments. The closed loop task, with its requirement for refined movements based on feedback, is more affected by frame time variations. This is perhaps to be expected since the feedback movement will be subject to the usual overshoots and corrections that one gets using feedback under varying conditions [5]. The more predictive open loop task tends to smooth out these variations, as long as they aren't too extreme.

Finally we see a significant effect due to the period of the frame time deviation at the longer frame time. Again this shows up mostly in the placement time (and marginally in the grab time) with performance being worse for the longer period oscillation than for the shorter one. Presumably this effect is due to the slower changes in frame time amplitude. (For example, more consecutive frames are spent at longer frame times.) In future studies it may be worthwhile to extend to even longer period oscillations, though, for the application designer and user, there is obviously a point of diminishing returns in extending to longer periods.

We have done other experiments [13], for the same set of tasks, that shed light on the study reported here. These experiments use a typical time series of frame time oscillations from a VR application. This time series is shifted and scaled to provide a set of different average frame rates and frame rate (rather than frame time) deviation amplitudes; thus the deviations are not as well controlled as in this study and the deviation periods are not well characterized. However, the experiments overlap the average frame rates used here. They show a similar trend in performance in going from lower to higher frame rates. Further, since deviations were more extreme and went to lower frame rates, the experiments show grab times can be affected at frame rates around 10 fps. Also, at higher frame rates (around 17 fps), the more extreme deviations (to lower frame rates) cause a significant effect on placement performance.

## 7 Conclusions and future work

In conclusion, this study provides a first careful analysis of the effects of frame time deviation amplitudes and periods on performance of typical VR tasks. The results show that at frame times (50 ms or 20 fps) in the range deemed acceptable for many applications, deviations up to 40% (of the average frame time) in amplitude and over a range of periods, do not affect task performance. This is important information for VR application designers. Precise, predictive algorithms are needed to keep frame time variations less than 10% for highly varied walkthrough environments [5], but feedback mechanisms [5,6], continuous level of detail methods with appropriately chosen parameters [9], or combination feedback/ predictive methods may be adequate much of the time if frame time consistency requirements are not so strict. Certainly virtual applications are moving towards significantly more complicated and larger environments that may include combinations of architectural elements, moving objects, high resolution terrain, dynamically added or removed objects, and simulated events. Managing these environments will be much more complicated than at present, and the tools may not give results that are so precise and predictable. In this situation, designers will want to know the range of acceptability for frame time fluctuations.

This study also provides new information to develop general understanding of the relationship between display variables and performance in a VE. Such information is always welcome because, compared say to window-based interfaces, VEs are significantly understudied via controlled experiments and significantly more complicated. In particular this study shows that to correctly predict performance, one must take into account not only average frame time but also the distribution and period around that mean, at least for certain ranges of frame times and fluctuations. With results such as these, one can eventually build up a design space from which to derive task-specific design principles.

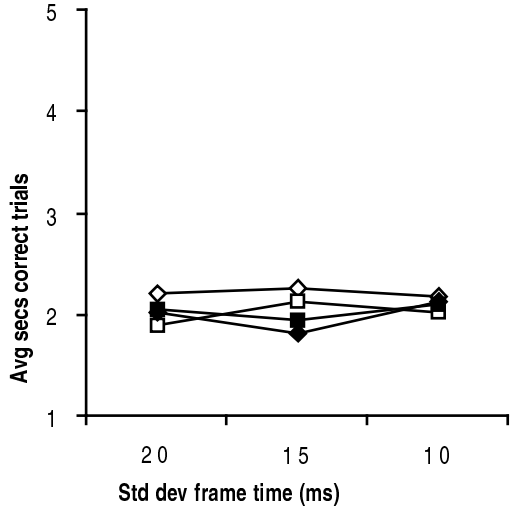

Figure 4b: Mean grab and place times for the mean frame time of 50 ms, with task type indicated by point color and frame time frequency by shape.

This work could be extended in several ways in the future. One could look at other tasks in the environment such as navigation involving "walking" or "flying", reaction time tasks, search tasks, and so on. Certainly the performance space should be filled in with studies at other frame times and fluctuation amplitudes. The studies begun in [13], looking at non-uniform variations in frame time or frame rate, could also be continued for other types of fluctuation patterns. Here it would be useful to come up with a measure of the fluctuation distribution so that one could classify the distributions in a quantitative way. Finally it would be useful to look at the effects of lag separately. These experiments vary lag as they vary frame time, but one could set up an environment with a fixed delay due to rendering and display and then vary the lag time. Since several researchers [2,10,14,15] say that lag is the dominant component affecting performance, a study of lag variations could be quite revealing.